\documentclass[prl,aps,superscriptaddress,preprint,floatfix]{revtex4-1}

\usepackage{graphicx}
\usepackage{verbatim}
\usepackage{mathrsfs}
\usepackage{amsmath,amsfonts,amssymb}
\usepackage{wrapfig}
\usepackage{graphicx}
\usepackage{bbm}
\usepackage{graphics}

\def\3{2.8in}
\def\2{2.5in}
\def\4{3.0in}

\def \beq {\begin{equation}}
\def \eeq {\end{equation}}
\begin{document}

\title{Observation of monolayer valence band spin-orbit effect and induced quantum well states in MoX$_{2}$}

\author{Nasser Alidoust}\affiliation {Joseph Henry Laboratory, Department of Physics, Princeton University, Princeton, New Jersey 08544, USA}

\author{Guang Bian}\affiliation {Joseph Henry Laboratory, Department of Physics, Princeton University, Princeton, New Jersey 08544, USA}

\author{Su-Yang Xu}\affiliation {Joseph Henry Laboratory, Department of Physics, Princeton University, Princeton, New Jersey 08544, USA}

\author{Raman Sankar} \affiliation{Center for Condensed Matter Sciences, National Taiwan University, Taipei 10617, Taiwan}

\author{Madhab Neupane}\affiliation {Joseph Henry Laboratory, Department of Physics, Princeton University, Princeton, New Jersey 08544, USA}

\author{Chang Liu}\affiliation {Joseph Henry Laboratory, Department of Physics, Princeton University, Princeton, New Jersey 08544, USA}

\author{Ilya Belopolski}\affiliation {Joseph Henry Laboratory, Department of Physics, Princeton University, Princeton, New Jersey 08544, USA}

\author{Dong-Xia Qu}\affiliation {Lawrence Livermore National Laboratory, Livermore, California 94550, USA}

\author{Jonathan D. Denlinger}\affiliation {Advanced Light Source, Lawrence Berkeley National~Laboratory, Berkeley, California 94305, USA}

\author{Fang-Cheng Chou} \affiliation{Center for Condensed Matter Sciences, National Taiwan University, Taipei 10617, Taiwan}

\author{M. Zahid Hasan}\affiliation {Joseph Henry Laboratory, Department of Physics, Princeton University, Princeton, New Jersey 08544, USA}


\begin{abstract}
\textbf{Transition metal dichalcogenides have attracted much attention recently due to their potential applications in spintronics and photonics as a result of the indirect to direct band gap transition and the emergence of the spin-valley coupling phenomenon upon moving from the bulk to monolayer limit. Here, we report high-resolution angle-resolved photoemission spectroscopy on MoSe$_{2}$ (molybdenum diselenide) single crystals and monolayer films of MoS$_{2}$ grown on Highly Ordered Pyrolytic Graphite substrate. Our experimental results, for the first time, resolve the two distinct bands at the Brillouin zone corner of the bulk MoSe$_{2}$, and provide evidence for the critically important spin-orbit split bands of the monolayer MoS$_{2}$. Moreover, by depositing potassium on cleaved surfaces of these materials, the process through which quantum well states form on the surfaces of transition metal dichalcogenides is systematically imaged. We present a theoretical model to account for the observed spin-orbit splitting and the rich spectrum of the quantum well states observed in our experiments. Our findings taken together provide important insights into future applications of transition metal dichalcogenides in nanoelectronics, spintronics, and photonics devices as they critically depend  on the spin-orbit physics of these materials.}
\end{abstract}
\date{\today}
\maketitle

In the recent years, with the discovery of two-dimensional states in graphene \cite{Novoselov2005, Zhang2005, Geim2007} and topological insulator surfaces \cite{Hasan2010, Qi2011, Hasan2011}, there has been an outburst of research activities to understand the physical properties of materials featuring two-dimensional (2D) states and their applications to optoelectronics and spintronics. One family of 2D materials that has recently attracted much attention is transition metal dichalcogenides (TMDCs) such as monolayers and few-layers of MX$_{2}$ (M = Mo, W and X = S, Se) \cite{Wang2012b, Balendhran2013}. These materials are easy to cleave, exfoliate, and fabricate \cite{Coleman2011, Castellanos-Gomez2012}, and have band gaps comparable in size to that of silicon (1.1 eV - 1.4 eV in bulk and 1.5 eV - 1.9 eV in monolayers) with the predicted property of transitioning from an indirect to a direct one upon reduction in size from bulk to monolayer  \cite{Wang2012b, Kobayashi1995, Lebegue2009, Ding2011, Kuc2011, Mak2010, Splendiani2010}, making them suitable for potential use in electronics applications such as thin film transistors. Furthermore, it has been predicted that the exotic spin-valley coupling phenomena occurs in monolayers of TMDCs, thus making them appropriate for spintronics and valleytronics devices \cite{Zhu2011, Xiao2012, Zeng2012, Zeng2013}. Nevertheless, even though many recent studies have examined the possibility of using transition metal dichalcogenides in single-layer transistors \cite{Radisavljevic2011, Yoon2011, Yin2012, Zhang2012}, and have investigated their optical properties and potential use in optoelectronics \cite{Frindt1963a, Frindt1963b, Kam1982, Zeng2012, Mak2010, Splendiani2010}, surprisingly there have only been a few experimental works on uncovering the electronic band structures of the bulk, few-layers, and monolayers of these materials and their modification under external effects such as doping, surface deposition, or heterostructuring. The few existing momentum-resolved spectroscopic studies on these materials also lack the resolution and depth to highlight the important features of their band structures \cite{Boker2001, Mahatha2012, Mahatha2012a, Mahatha2013, Jin2013}. For instance, only recently using micro-ARPES the evolution of the valence band at the corner of the Brillouin zone (BZ) from higher to lower binding energies upon reducing the number of layers to monolayer has been demonstrated \cite{Jin2013}.

The interest in monolayers of the MX$_{2}$ series lies in the emergence of the direct band gap as a consequence of the disappearance of van der Waals interlayer interactions in the  monolayer limit \cite{Splendiani2010, Wang2012b}, as well as the valence band strong spin-orbit splitting due to the loss of inversion symmetry \cite{Zhu2011, Xiao2012, Wang2012b, Jin2013}. However, since films of TMDCs on typical substrates such as Si/SiO$_{2}$ grow in very small flakes of $\sim$ 10 $\mu$m $\times$ 10 $\mu$m and are usually composed of islands of variable thicknesses next to each other, tracking these two properties in a momentum-resolved fashion has proven to be challenging. Furthermore, the splitting of the valence band top into two distinct degenerate bands in the bulk limit due to both strong spin-orbit coupling and interlayer hopping \cite{Zahid2013} could be of interest and has not been resolved to date. Moreover, until very recently MoS$_{2}$ has experimentally gotten the most attention amongst various members of the TMDCs family such as MoSe$_{2}$, WS$_{2}$, and WSe$_{2}$. This is despite the fact that the stronger spin-orbit coupling of the later compounds make them more attractive for spintronics applications.

Thus, in this work we utilize angle-resolved photoemission spectroscopy (ARPES) to directly map the electronic band structure of bulk MoSe$_{2}$ across the entire BZ. We then study the evolution of its band structure upon  \textit{in-situ} surface deposition with an alkali metal (potassium). Our results demonstrate the formation of a nearly free 2D electron gas (2DEG) within the potassium overlayers confined to the surface of MoSe$_{2}$ in the form of quantum well states (QWSs). This quantum confinement of potassium layers could potentially enhance the optoelectronics performance of bulk and thin films of the TMDCs and be utilized in optoelectronics applications of these materials. Moreover, we resolve the band structure of a monolayer of MoS$_{2}$ grown on Highly Ordered Pyrolytic Graphite (HOPG) substrate using the chemical vapor deposition (CVD) growth method and find evidence suggesting the existence of the split valence band induced by strong spin-orbit coupling and broken inversion symmetry. We compare our ARPES results with first-principles theoretical calculations to gain microscopic insights on the electron behavior in these materials which is critically important for optimizing their potential performance in device settings.

\bigskip
\textbf{Results}

\textbf{Crystal structure of MX$_{2}$.} The crystal structure of MX$_{2}$ (M = Mo, W and X = S, Se) is a layered structure with layers of M atoms sandwiched between two layers of X atoms. These layers, which are held together through van der Waals forces, are shifted relative to each other in such a way that M atoms in one layer are placed directly above the X atoms in the neighboring layers \cite{Wang2012b, Balendhran2013, Mahatha2012}. This crystal structure and the corresponding Brillouin zone are shown in Figs. 1\textbf{b} and 1\textbf{c}, respectively. The center of the BZ is the $\Gamma$ point, with $M$ points the center of the hexagonal BZ edges, and $K$ points the corners of the hexagon. $\overline{\Gamma}$, $\overline{M}$, and $\overline{K}$ are the projection of these points onto the 2D surface BZ as illustrated in Fig. 1\textbf{c}. X-ray photoemission spectroscopy (XPS) measurements of the single crystals of MoSe$_{2}$ used in our studies reveal sharp peaks corresponding to Mo 3$s$, 3$p$, and 3$d$ core levels, as well as those of Se 3$p$ and 3$d$ as shown in Fig. 1\textbf{a}, thus indicating the high quality of the studied samples.

\bigskip
\textbf{Electronic band structure of MoSe$_{2}$.} We start by investigating the band structure of MoSe$_{2}$. The $k$ - $E$ maps of the electronic band structure measured by ARPES (left panels in Figs. 2\textbf{a} and 2\textbf{b}) along the two high symmetry  directions of the BZ, $\Gamma - M$ and $\Gamma-K$, show extraordinary resemblance to the calculated bands from the first-principles calculations (right panels in Figs. 2\textbf{a} and 2\textbf{b}). The ARPES measurements here were performed at a photon energy of $h\nu$ = 90 eV. The kinetic energy value of the Fermi level is determined by measuring that of the polycrystalline gold separately at the same photon energies and under the same experimental conditions. The valence band maximum (VBM) is found to be located at the $\Gamma$ point of the BZ at a binding energy of 1.2 eV. The valence band just below the topmost band at binding energies of $\sim$ 1.7 eV - 1.9 eV, which appears clearly in our first-principles calculations has a somewhat weaker intensity in our ARPES measurements due to matrix elements effect.

One important feature of our measurements is the unambiguous observation of the two distinct bands near the top of the valence band along the $\Gamma-K$ direction of the BZ. These bands are highlighted in Fig. 2\textbf{c}, a zoomed-in version of the top part of the valence band along the $\Gamma-K$ direction, and are well in agreement with band structure calculations conducted here as well as the ones reported in earlier studies \cite{Wang2012b, Lebegue2009, Ding2011}. Constant binding energy contours at representative binding energies of 1.4 eV, 1.7 eV, and 2.0 eV are presented in Figs. 2\textbf{d} - 2\textbf{f}. At $E_{\text{B}} = 1.7$ eV, there exists only one electron-like pocket around each of the $K$ points of the BZ (see Fig. 2\textbf{e}), whereas two concentric pockets are observed at $E_{\text{B}} = 2.0$ eV (see Fig. 2\textbf{f}), clearly resolving the two distinct bands at the BZ corners.

\bigskip
\textbf{Spin-orbit induced splitting of the valence band in monolayer TMDCs.} Fig. 1\textbf{d} shows our first-principles band structure calculations for bulk (top row) and monolayer (bottom row) MoS$_{2}$, with (right column) and without (left column) considering spin-orbit coupling. These demonstrate the indirect to direct band gap transition on going to the monolayer limit and the spin-orbit splitting of the monolayer valence band in TMDCs. The Fermi level in these $k$ - $E$ maps is indicated by the dashed lines. In the next step we turn our attention to monolayer MoS$_{2}$ on HOPG substrate. This conducting substrate was deliberately chosen to avoid the electronic charging of the film during the photoemission process. The $k$ - $E$ map of the electronic band structure of this sample along the high symmetry direction $\overline{\Gamma}-\overline{K}$ reveals the electronic bands of monolayer MoS$_{2}$ as well as the linear $\pi$ bands of graphite from the HOPG substrate at the $\overline{K}$ point of the BZ. We have conducted first-principles calculations of monolayer MoS$_{2}$ (right panel in Fig. 3\textbf{b}) and the hybrid structure of this monolayer and graphite (left panel in Fig. 3\textbf{b}), which indicate the general agreement with our ARPES findings. Despite the existence of the graphite's $\pi$ bands at the $\overline{K}$ points, and their merging with the MoS$_{2}$ valence band around these points of the BZ, we can still resolve a large section of the MoS$_{2}$ valence band top near $\overline{K}$ along the high symmetry direction $\overline{\Gamma}-\overline{K}$, as highlighted in Fig. 3\textbf{c}, a zoomed-in version of Fig. 3\textbf{a} near the top of the valence band.

Moreover, we note the broadening of the valence band close to the Fermi level along the  $\overline{\Gamma}-\overline{K}$ high symmetry direction (see Fig. 3\textbf{c}). Performing 2D curvature analysis \cite{Zhang2011} of the obtained bands, we resolve two split branches of the valence band of MoS$_{2}$ along this high symmetry direction (close to the $\overline{K}$ point) as shown in Fig. 3\textbf{d}. The white dotted lines are guides to the eyes for these two spin-orbit split branches of the valence band. This is consistent with the spin splitting scenario predicted to exist in this material. Monolayers of TMDCs lack inversion symmetry and are non-centrosymmetric. The in-plane confinement of electron motion and high mass of the elements in these materials result in strong spin-orbit splitting of the valence band. This property makes these materials promising candidates for applications in spintronics devices \cite{Wang2012b, Zhu2011, Xiao2012, Zeng2013}. Our observation here is the first momentum-resolved experimental evidence for such spin-orbit splitting. Further spin-resolved ARPES studies are needed to identify the spin configuration of these split bands.

To further demonstrate the orbital property and spin texture of monolayer MoS$_{2}$, we calculate the charge densities and spin expectation values of the lowest two conduction states and the topmost two valence states at $\overline{K}$ marked in Fig. 1\textbf{d} as C1, C2, V1, and V2, respectively. The charge density plots shown in Fig. 3\textbf{e} indicate that the dominant orbital components are Mo $4d_{3z^{2}-r^{2}}$ and Mo $4d_{x^{2}-y^{2}} + 4d_{xy}$  for C1(C2) and V1(V2), respectively. The calculated spin expectation values of states near $\overline{K}$ show nearly out-of-plane spin polarization, which can be attributed to an interplay of the pure in-plane electron motion and the inversion symmetry breaking crystal potential. The out-of-plane character of Mo $4d_{3z^{2}-r^{2}}$ explains the negligible spin splitting between the C1 and C2 states.

On the other hand, the calculated energy splitting between V1 and V2 states is 138 meV at $\overline{K}$, as a consequence of the prominent in-plane charge density variation of Mo $4d_{x^{2}-y^{2}} + 4d_{xy}$ orbital, which is in good agreement with previous results from Density Functional Theory calculations \cite{Zhu2011}. According to the calculated spin expectation values, we deploy a straightforward method to determine the spin orientation of the V1 and V2 states, which are important when considering the selection rule of polarized photon excitations. Viewing monolayer MoS$_{2}$ from top, since sulfur atoms sit on the hexagonal lattices it's alway possible to find a primitive vector connecting to adjacent sulfur atoms in the same direction as $\overline{\Gamma}-\overline{K}$. Call this vector \textbf{a} and drawing another vector \textbf{b} from the sulfur atom at the origin of \textbf{a} to the nearest Mo atom, the spin orientation of V2 is in the direction of $\mathbf{a} \times \mathbf{b}$ and that of V1 points to the opposite direction.

\bigskip
\textbf{Observation of QWSs of a nearly free 2DEG by surface deposition.} In order to study the effect of alkali metal deposition on the surface of bulk MoSe$_{2}$, we deposit potassium (K) \textit{in-situ} on cleaved surfaces of  crystals of this material, and track the evolution of the band structure spectra as the deposition occurs. Deposition was performed at $T$ = 60 K with an estimated deposition rate of $\sim$ 0.08 - 0.09 $\text{\AA}$min$^{-1}$. Fig. 4\textbf{a} shows the evolution of the band structure $k$ - $E$ map along the $\Gamma-M$ direction of the BZ during the deposition process. We observe that the VBM at $\Gamma$ moves closer to the Fermi level by about 200 meV. We associate this shift with the formation of a conducting layer on the surface of our sample after potassium deposition. More specifically, the VBM of the MoSe$_{2}$ sample is at $E_{\text{B}} \sim 1.0$ eV, but before deposition is shifted down by about 200 meV due to the electronic charging effect. Next, we observe the formation of new electronic states in the band gap of MoSe$_{2}$, extending from the Fermi level to the valence band (see middle panels in Fig. 4\textbf{a}). Upon the completion of the deposition process these in-gap states are completely formed as shown in the last panel of Fig. 4\textbf{a} and Fig. 4\textbf{b}.

We note the formation of electron-like pockets on the Fermi surface after the completion of the deposition process. These pockets appear right at the $K$ points of the BZ. Since the energy spacing between these pockets and the VBM ($\sim 0.8$ eV) is smaller than the indirect gap of MoSe$_{2}$ ($\sim 1.1$ eV) we rule out the possibility of these pockets being a part of the conduction band of MoSe$_{2}$. The resulting Fermi surface after potassium deposition shows two concentric circular shaped pockets around the center of the BZ, as well as the aforementioned pockets at the $K$ points (Fig. 4\textbf{d}). The circular pockets around $\Gamma$ are identified as the first and second QWSs, whereas the pockets at $K$ as resulting from the surface states of potassium. We furthermore conduct first-principles calculations of potassium layers formed on the surface of bulk MoSe$_{2}$ and find that our experimental observation is in qualitative agreement with two atomic monolayers of potassium deposited on the surface of MoSe$_{2}$, as shown in Fig. 4\textbf{c}. The possibility of directly accessing the conduction band of bulk and monolayer TMDCs through intercalation and adsorption of alkali metals has been theoretically predicted \cite{Dolui2013}. Our results suggest that adsorption of alkali metals on the surface of bulk MoSe$_{2}$ results in the formation of QWSs and a nearly free 2DEG rather than electron doping and providing access to the conduction band.

\bigskip
\textbf{Discussion}

The splitting of the valence band along the $\overline{\Gamma}-\overline{K}$ high symmetry direction of the monolayer MoS$_{2}$ and other TMDCs which is due to strong spin-orbit coupling (because of the high mass of the elements) and lack of inversion symmetry, has been suggested in previous theoretical studies as well as our first-principles calculations here \cite{Wang2012b, Zhu2011}. For the bulk, this splitting is caused by the combination of interlayer interaction and spin-orbit coupling. In this case, there is no spin splitting between these two bands and they are indeed spin-degenerate (Kramers' degeneracy) due to the presence of inversion symmetry along with time reversal symmetry. But, the strong spin-orbit coupling in TMDCs results in enhanced splitting of these spin-degenerate bands at the $K$ point of the BZ (see Fig. 1\textbf{d}) \cite{Zahid2013}. Our results resolve these two distinct bands at the corners of the BZ through ARPES measurements and theoretical first-principles calculations. Furthermore, we have provided the first momentum-resolved experimental evidence of the spin-split valence band of monolayer MoS$_{2}$ grown on HOPG substrate. However, further spin-resolved ARPES measurements are needed to confirm the spin configuration of these two split bands.

Recently, experimental evidence and theoretical prediction of valley confinement has been suggested in monolayer MoS$_{2}$. Specifically, it has been suggested that the split valence band at neighboring $K$ points of the BZ ($K$ and $K^{\prime}$) posses opposite out of plane spin configurations \cite{Xiao2012, Zeng2012, Zeng2013}. The spin-orbit splitting observed in our study of monolayers of MoS$_{2}$ paves the way to more detailed studies of spin and valley physics in TMDCs. It also serves as a guide for spintronics and valleytronics applications of TMDCs.

Our results also track the evolution of the band structure of MoSe$_{2}$ throughout the surface deposition process with an alkali metal (potassium). We note that short after the deposition process has started potassium atoms sit on the surface of MoSe$_{2}$ in a manner that results to the partial formation of the first QWS (see second panel in Fig. 4\textbf{a}). Consequently, the first QWS appears in full, as shown in the third panel of Fig. 4\textbf{a}, and finally the first two QWSs are completely formed. In the process of surface deposition, 2D conduction channels (quantum well states) emerge on the surface of the semiconducting MoSe$_{2}$ (with an indirect insulating energy gap of $\sim$ 1.8 eV) and, as a consequence, the electronic behavior of the system changes dramatically. This tunable feature can be utilized to quantitatively control the electronic properties of this quasi-2D material. Our investigation of the dynamical process in which a nearly free 2DEG is formed on TMDCs could have important implications for future optoelectronics studies and applications of TMDCs using surface manipulations by elements such as alkali metals on bulk, few-layers, and monolayers of these materials.

In conclusion, we have conducted detailed high-resolution ARPES and first-principles calculations studies of the electronic structure of both bulk and monolayer TMDCs. We have resolved the two distinct spin-degenerate valence bands of the bulk MoSe$_{2}$, and have provided evidence for the spin-orbit split bands of the monolayer MoS$_{2}$, at the corners of the BZ. We have furthermore investigated the process of the formation of QWSs on the surface of bulk MoSe$_{2}$, and demonstrated the formation of a 2DEG on the cleaved surfaces of TMDCs. Our findings shed light on future applications of bulk and monolayer MoX$_{2}$ in electronics and spintronics devices.

\bigskip
\textbf{Methods}

\textbf{Electronic structure measurements.} ARPES measurements were performed with incident photon energies of 35 - 100 eV at beamline 4.0.3 of the Advanced Light Source (ALS) in the Lawrence Berkeley National Laboratory (LBNL). Samples were cleaved \textit{in-situ} between 10 and 20 K at chamber pressure better than $5\times10^{-11}$ torr resulting in shiny surfaces. Energy resolution was better than 15 meV and momentum resolution was better than 1\% of the surface BZ.

\textbf{Sample growth.} To prepare MoS$_{2}$ films, molybdenum chloride (MoCl5) and sulfur powder were used as precursors and the growth of monolayer films was carried out in a CVD chamber. The ultrasonicated cleaned HOPG substrates were loaded into the CVD chamber 6 cm away from the  central hot zone. Argon (Ar) gas was used as carrier gas in a pressure of 3 torr. Single crystals of MoSe$_{2}$ were grown by the chemical vapor transport method, using I$_{2}$ as the transporting agent. They were formed in silver-colored, graphite-like, single crystalline platelets up to 10 $\times$ 10 mm$^{2}$ in surface area and 2 mm in thickness (inset in Fig. 1\textbf{a}).

\textbf{First-principles calculation methods.} First-principles calculations of the electronic structures were performed using HGH-type pseudo-potentials \cite{Hartwigsen1998} and a plane-wave basis set. The main program employed was developed by the ABINIT group \cite{Gonze2002, Gonze2005}. Spin-orbit coupling was included, where appropriate, using the relativistic LDA approximation. The experimental resolution was taken into account in the spectral simulations of Figs. 2, 3, and 4, but the matrix elements effect was not considered.



\begin{figure*}[H]
\centering
\includegraphics[width=14cm]{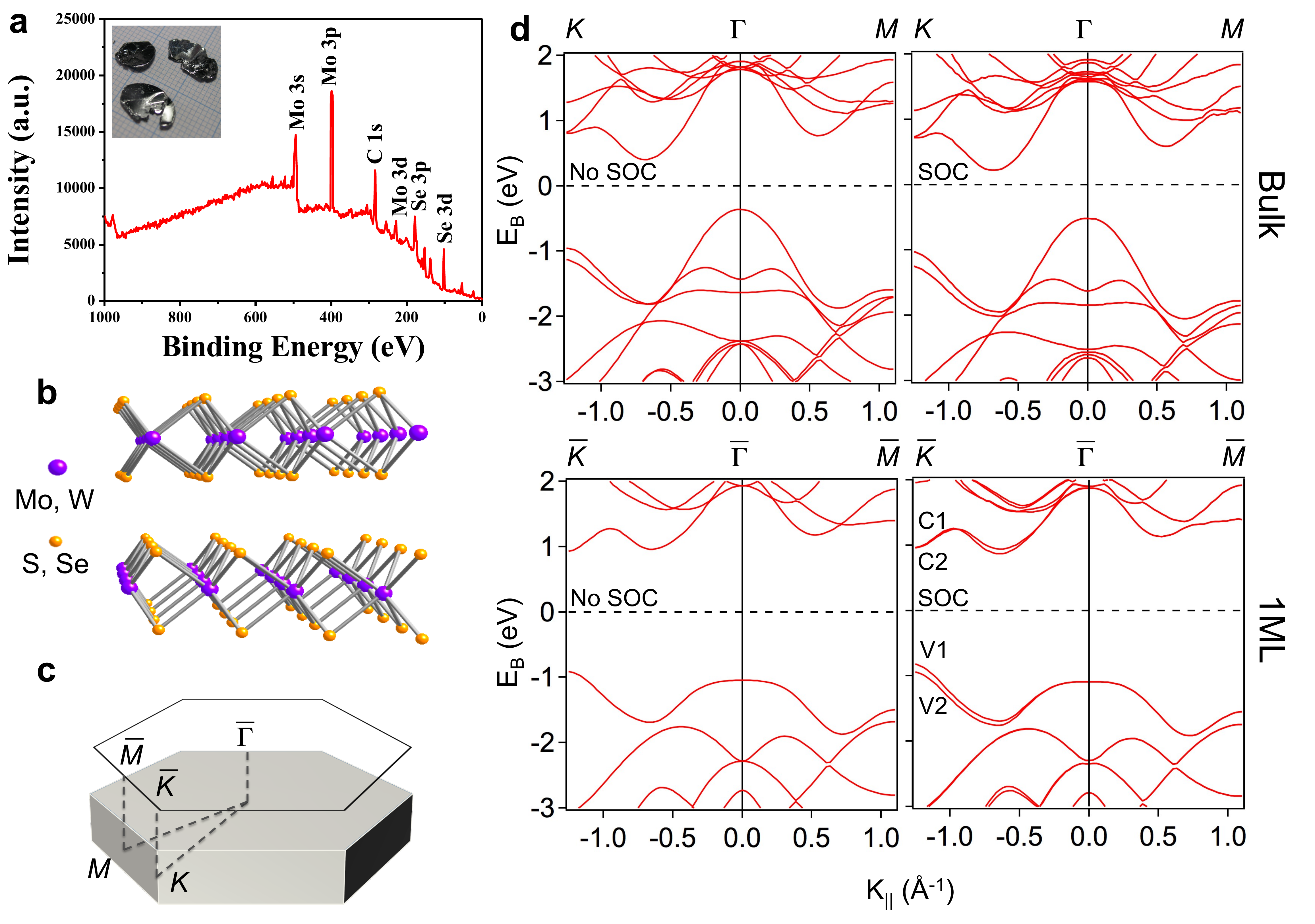}
\caption{\textbf{Structure and characterization of TMDC materials}.  \textbf{a}, XPS measurement of the studied MoSe$_{2}$ crystals, showing both Mo and Se peaks. The inset shows a photograph of these crystals. \textbf{b}, Three dimensional crystal structure of layered TMDCs, with the chalcogen atoms (X) in orange and the metal atoms (M) in purple. \textbf{c}, The corresponding bulk and surface BZ with the high symmetry points marked. \textbf{d}, First-principles band structure calculations for bulk (top row) and monolayer (bottom row) MoS$_{2}$ with (right column) and without (left column) considering spin-orbit coupling, demonstrating the indirect to direct band gap transition on going to the monolayer limit and the spin-orbit splitting of the monolayer valence band. The Fermi level is indicated by the dashed lines.}
\end{figure*}

\begin{figure*}
\centering
\includegraphics[width=16.5cm]{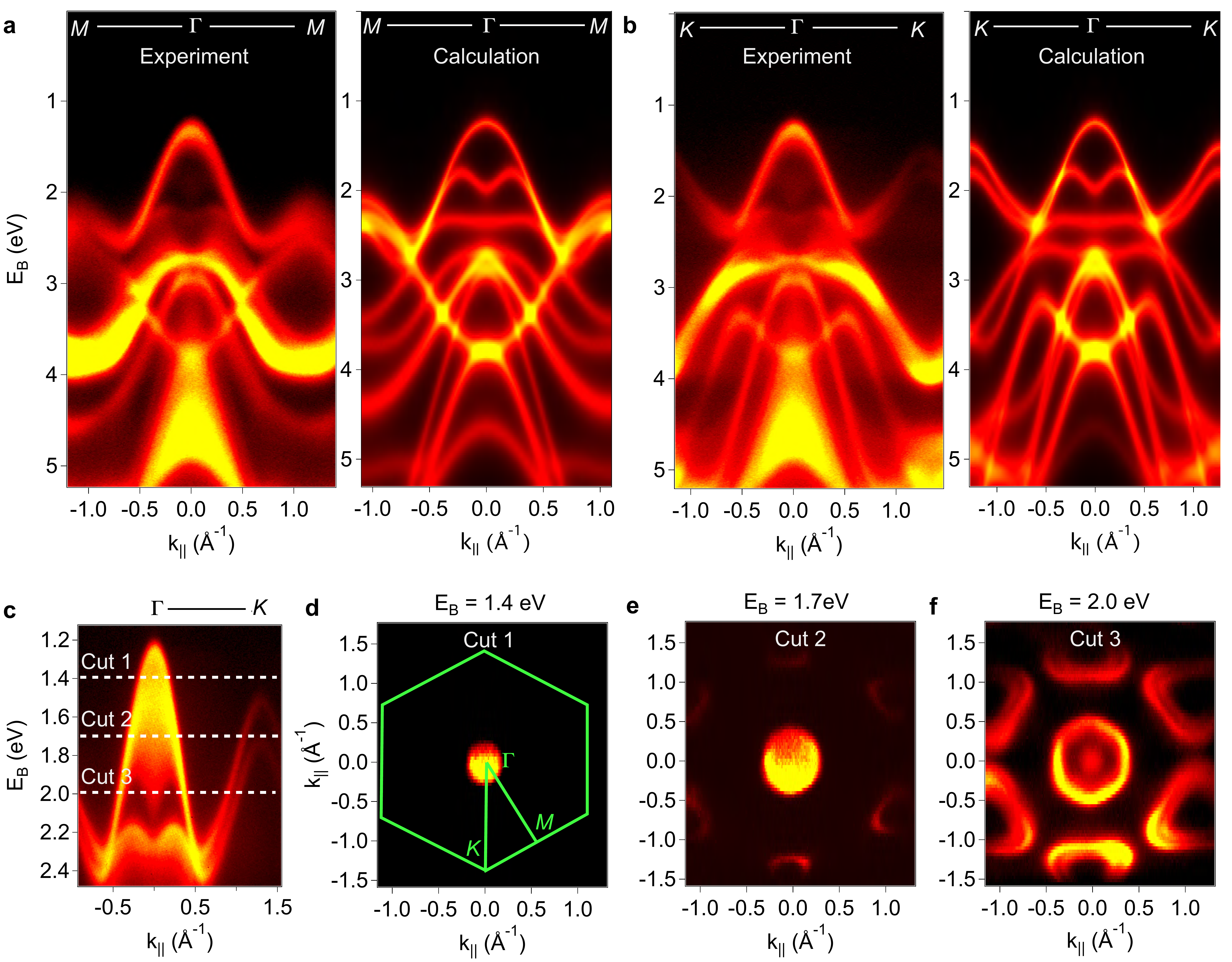}
\caption{\textbf{Electronic band structure and spin-degenerate bands of bulk MoSe$_{2}$.} \textbf{a}, ARPES electronic structure measurements (left panel) and first-principles calculated electronic structure (right panel) of MoSe$_{2}$ crystals along the high symmetry direction $\Gamma-M$. \textbf{b}, Same as \textbf{a} for measurements and calculations along the high symmetry direction $\Gamma-K$. \textbf{c}, Zoomed-in version of the ARPES electronic structure spectra along the $\Gamma-K$ direction near the top of the valence band, highlighting the two distinct spin-degenerate bands at the $K$ point of the BZ. \textbf{d}, \textbf{e}, \textbf{f}, Constant binding energy contours at $E_{\text{B}}$ = 1.4 eV, $E_{\text{B}}$ = 1.7 eV, and $E_{\text{B}}$  = 2.0 eV which are indicated in panel \textbf{c}. The green hexagon in \textbf{d} represents the first BZ. The two distinct bands can also be distinguished in the contour at $E_{\text{B}}$  = 2.0 eV in \textbf{f} from the electron-like pockets around the $K$ points. The ARPES measurements were performed at photon energy of $h\nu$ = 90 eV and temperature of $T$ = 20 K.}
\end{figure*}

\begin{figure*}
\centering
\includegraphics[width=16cm]{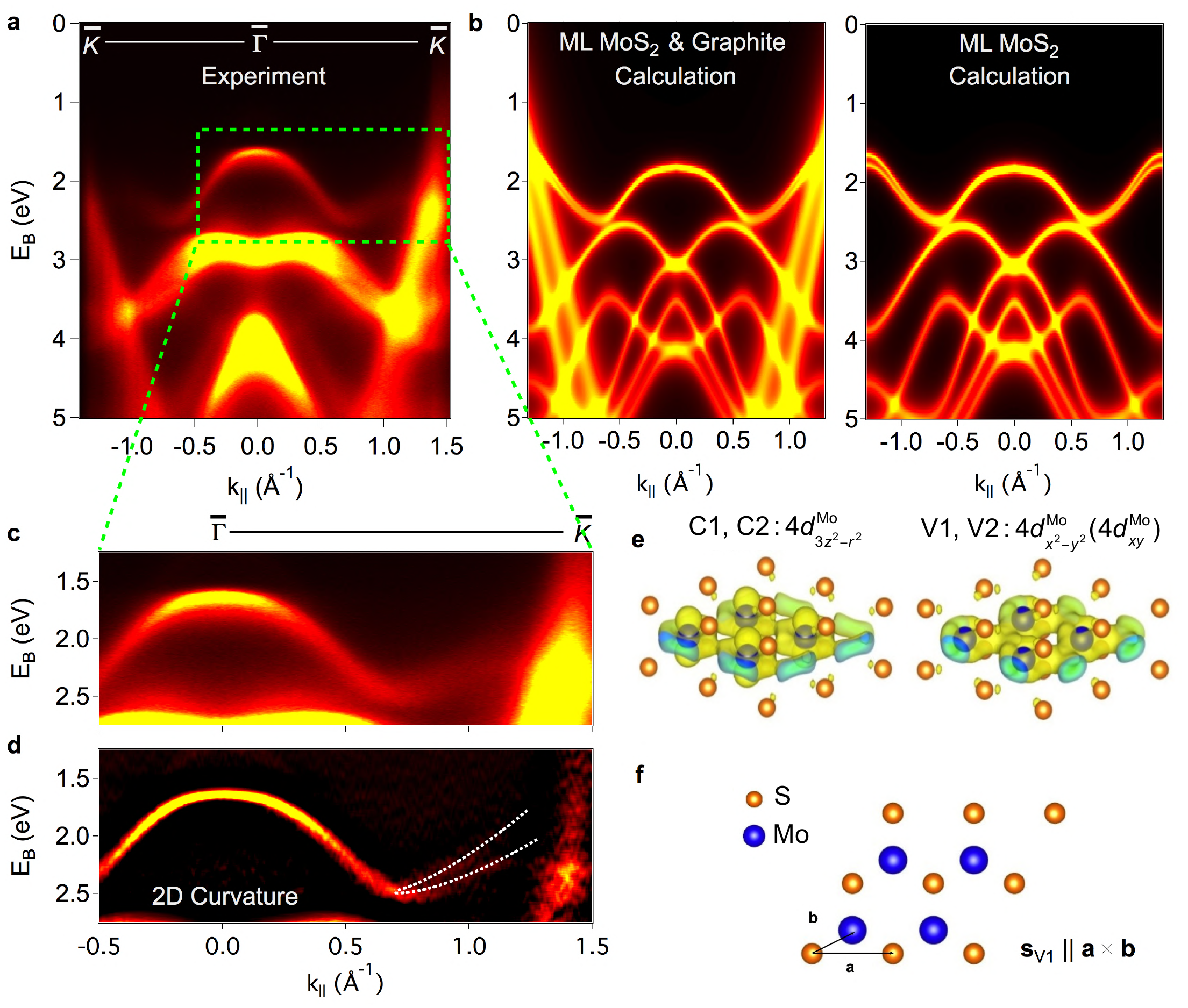}
\caption{\textbf{Spin-orbit induced splitting of the valence band in monolayer MoS$_{2}$.} \textbf{a}, ARPES electronic structure measurements of monolayer MoS$_{2}$ on HOPG substrate along the high symmetry direction $\overline{\Gamma}-\overline{K}$, showing the observation of MoS$_{2}$ monolayer electronic bands as well as the $\pi$ bands of graphite at the $K$ point of the BZ. \textbf{b}, First-principles calculated electronic structure measurements of monolayer MoS$_{2}$ (right panel) and monolayer MoS$_{2}$ accompanied by graphite (left panel) along the high symmetry direction $\overline{\Gamma}-\overline{K}$. \textbf{c}, Zoomed-in version of the ARPES electronic structure spectra along the $\overline{\Gamma}-\overline{K}$ direction. \textbf{d}, 2D curvature plot of the bands near the valence band top, showing the experimental observation of the predicted spin-orbit split valence band along $\overline{\Gamma}-\overline{K}$. The white dotted lines are guides to the eyes for these two spin-orbit split branches of the valence band. Here, the ARPES measurements were performed at photon energy of $h\nu$ = 110 eV and temperature of $T$ = 200 K. \textbf{e}, Charge densities of conduction and valence states at $\overline{K}$. \textbf{f}, Top view of monolayer MoS$_{2}$. The spin polarization of the V1 band is in the direction of $\mathbf{a} \times \mathbf{b}$, provided that $\mathbf{a}$ is aligned along the $\overline{\Gamma}-\overline{K}$ direction.}
\end{figure*}

\begin{figure*}
\centering
\includegraphics[width=16.5cm]{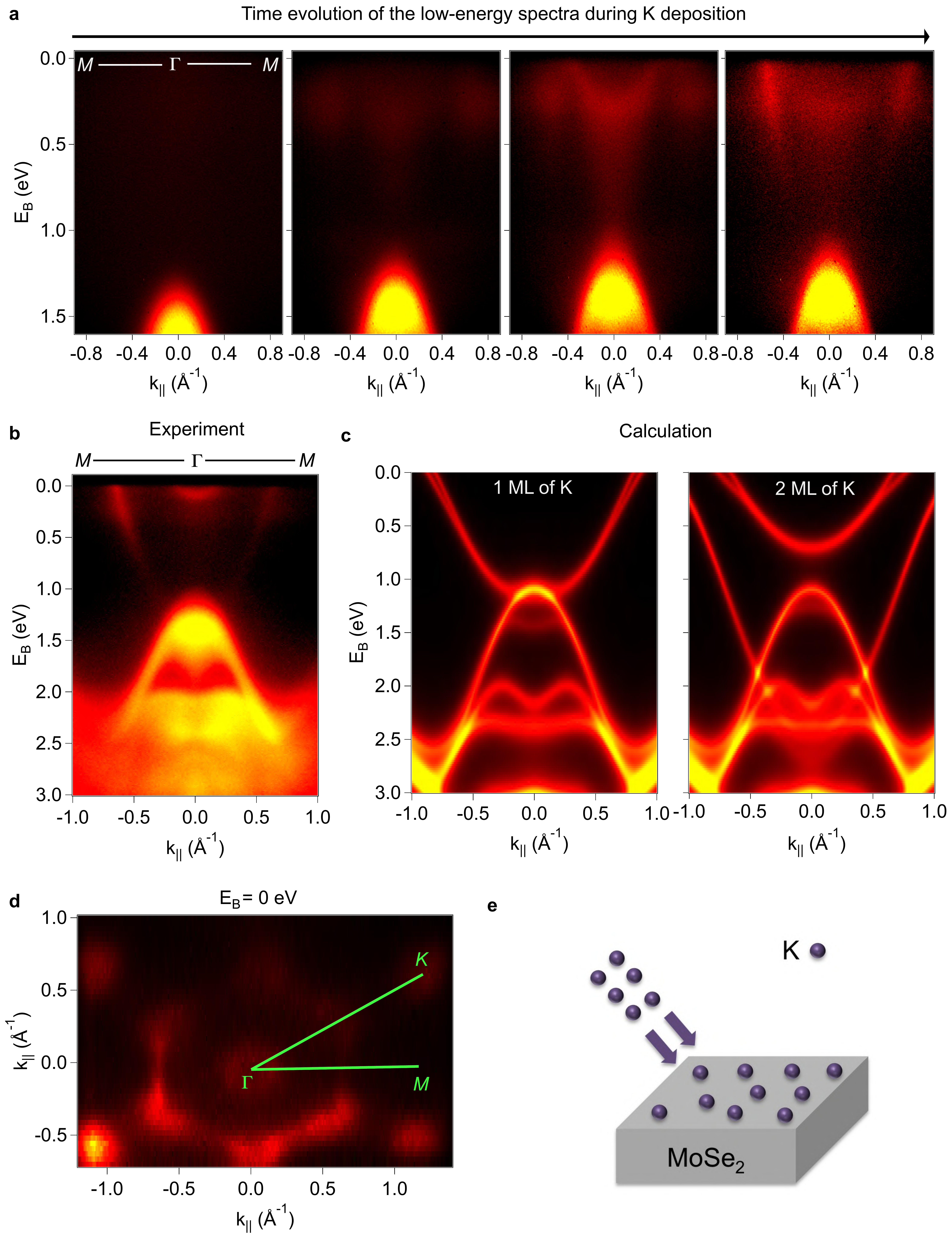}
\caption{\textbf{QWSs on TMDCs through surface potassium deposition.} \textbf{a}, Evolution of the low energy electronic structure of MoSe$_{2}$ during \textit{in-situ} surface potassium deposition as measured by}
\end{figure*}
\addtocounter{figure}{-1}
\begin{figure*}[t!]
\caption{ cont'd. ARPES, showing the process of formation of in-gap quantum well states. \textbf{b}, ARPES band structure spectra upon the completion of the potassium deposition process. \textbf{c}, First-principles calculated band structure of one and two atomic layers of potassium on the surface of bulk MoSe$_{2}$, indicating the formation of two QWSs in the ARPES measurements. \textbf{d}, The Fermi surface obtained after the potassium deposition process is completed, featuring the observation of two QWSs in the form of two electron-like circular pockets around $\Gamma$, as well as the surface states of potassium at $K$ points. The green lines represent the $\Gamma-M$ and $\Gamma-K$ directions of the BZ. Here, the ARPES measurements were performed at photon energy of $h\nu$ = 70 eV and temperature of $T$ = 60 K. \textbf{e}, Illustration of the \textit{in-situ} surface potassium deposition.}
\end{figure*}

\end{document}